\documentclass[journal]{IEEEtran}
 \newcommand{\ud}{\,\mathrm{d}}

\usepackage{amssymb, amsmath,rangecite}

 \usepackage[dvips]{color} 
\usepackage{lscape,longtable,nomencl}
\input{psfig.sty}
 \usepackage{epsfig,subfigure}
\usepackage[usenames,dvipsnames]{pstricks}
 
\newcommand{\evec}{{\bf{e}}}

\newcommand{\wvec}{{\bf{w}}}

\newcommand{\vvec}{{\bf{v}}}

\newcommand{\Hmat}{{\bf{H}}}

\newcommand{\Imat}{{\bf{I}}}

\newcommand{\Rmat}{{\bf{R}}}

\newcommand{\define}{\stackrel{\triangle}{=}}

\def\bnu{{\mbox{\boldmath $\nu$}}}

\def\bnu{{\mbox{\boldmath $\nu$}}}

\def\thetavec{{\mbox{\boldmath $\theta$}}}

\def\thetavecsmall{{\mbox{\boldmath {\scriptsize $\theta$}}}}

\newcommand{\be}{\begin{equation}}
\newcommand{\ee}{\end{equation}}
\newcommand{\beqna}{\begin{eqnarray}}
\newcommand{\eeqna}{\end{eqnarray}}

\title{PMU based Detection of Imbalance in Three-Phase Power Systems}
\author{Tirza~Routtenberg,~\IEEEmembership{Member,~IEEE},
        Yao~Xie,~\IEEEmembership{Member,~IEEE},
				Rebecca~M.~Willett,~\IEEEmembership{Senior~Member,~IEEE}
        and~Lang~Tong,~\IEEEmembership{Fellow,~IEEE}
\thanks{T. Routtenberg and L. Tong are with School of Electrical and Computer Engineering, Cornell University, Ithaca, NY 14853, United States, Email: \{tsr43,lt35\}@cornell.edu.}
\thanks{ Y. Xie is with the Milton Stewart School of Industrial and Systems Engineering, Georgia Institute of Technology, Email: yao.xie@isye.gatech.edu,}
\thanks{R. M. Willett is with the Department of Electrical and Computer Engineering, University of Wisconsin-Madison,
 Email: rmwillett@wisc.edu.}
\thanks{The work of T. Routtenberg and L. Tong is supported in part by the National Science Foundation under
Grant CNS-1135844.}
\vspace{-0.4cm}}

\begin{document}	\maketitle
 \begin{abstract}
The problem of imbalance detection in a  three-phase power system using a phasor measurement unit (PMU) is considered.
  A general model for the zero, positive, and negative sequences from a PMU measurement at off-nominal frequencies is presented and a hypothesis testing framework is formulated. 
	The new formulation takes into account  the fact that
	minor degree of imbalance in the system is acceptable and
	does not indicate subsequent interruptions, failures, or degradation of physical components.
	A generalized likelihood ratio test (GLRT) is developed  and  shown to be a function of the negative-sequence phasor estimator and the
		acceptable level of imbalances for nominal system operations.
		As a by-product to the proposed {\em{detection}} method, a constrained {\em{estimation}} of the positive and negative phasors and the frequency deviation is obtained for both balanced and unbalanced situations.
				The theoretical and numerical performance analyses  show improved performance over benchmark techniques and  robustness to the presence of additional harmonics.
\end{abstract}
\section{Introduction}
\label{sec:intro}
The three-phase power system is designed to operate at a nominal frequency in a near-balanced fashion \cite{PMUbook}.
  In practice,  frequency deviation and load imbalance are the norm rather than the exception. 
	 According to the American National Standards Institute (ANSI) report  \cite{ANSI_C84.1},
$2\%$ of the electrical distribution systems in USA have  a significant undesirable degree of imbalance,
leading to several serious consequences. First, frequency deviations and imbalances
 may be a precursor to more serious contingencies leading to possible blackouts \cite{Abur2002}, \cite{Jouanne_Banerjee_2001}.
In addition, substantial power imbalance causes excessive losses, overheating, insulation degradation,  a reduced lifespan of motors and transformers, and  interruptions in production processes 
{\cite{power_sys_book,Woolley_Milanovic_2012,BinaKashefi2011,Chen2013}}.

Thus, 
the ability to detect potentially harmful levels of imbalance   in various power systems is highly desirable  for the benefit of both the utility and customer 
 \cite{Jouanne_Banerjee_2001}, \cite{Woolley_Milanovic_2012}.
 However, in most phasor measurement applications it is common and acceptable  \cite{ANSI_C84.1}
 to have some degree of imbalance in the system  due to unbalanced loads and untransposed transmission lines \cite{PMUbook}.
To this end, effective algorithms and sophisticated methods are crucial for estimating  frequency deviations and phasors in the event of system imbalance and  detecting   an {\em{abnormal}} level of imbalance.
It is in this context that modern sensing devices, such as phasor measurement units (PMUs),  
  have the potential to provide rapid detection of contingencies and situational awareness
(see \cite{PMUbook} and references therein).

\vspace{-0.15cm}
\subsection{Summary of results}
In this paper we consider detection of voltage imbalance in three-phase power systems using the
native frequency output of PMU.
The contribution of the paper is threefold.
  First, we develop a statistical model that captures characteristics of imbalance from PMU output.  In particular, we provide the noise statistics and  demonstrate that,
 for a {\em{perfectly}} balanced power system, the PMU output is a single complex sinusoid, whereas, under imbalance,  the symmetrical components at the PMU output have two related frequencies.
The statistical model indicates that, at the nominal frequency,  imbalance is undetectable by using only the 
positive-sequence.   Therefore,  detection of imbalance   should be carried out by using   the negative-sequence and/or the zero-sequence in addition to the positive-sequence.
Second, we derive a hypothesis testing technique based on the principle of generalized likelihood ratio test (GLRT).   The proposed GLRT uses the constrained maximum-likelihood (CML) estimators of the 
frequency deviation and the three symmetrical component phasors under the balanced and unbalanced
 system operating conditions. 
Third, we analyze the performance of the proposed GLRT and provide simulation results in practical settings. 
 In particular, in Section \ref{theoretical_sec}, we present an analysis of false alarm probability from which we obtain a practical way of setting detection thresholds for given false alarm probabilities.  Simulations studies are presented in Section  \ref{simulation_sec}, where we demonstrate
the performance of the proposed GLRT for single-phase magnitude and phase imbalances.
We demonstrate that at high signal-to-noise ratios (SNR),
 the GLRT with estimated frequency coincides with the known-frequency GLRT and thus, the frequency estimation has no impact in this region.  
 Of particular importance is the evaluation of the robustness of the proposed algorithm in the presence of higher-order harmonics.
  We show that the probability of detection of the GLRT 
	for  non-sinusoidal  voltages  
 is close to those of the sinusoidal case.
 Therefore, the proposed
method can  also be used  in the
presence of inter-harmonics.  

\vspace{-0.15cm}
\subsection{Related works}
Under perfectly balanced three-phase operating conditions, the zero and negative sequences are absent,
 hence  the state-estimation and signal analysis in this case are carried out using only the positive-sequence model {\cite{Abur2002,Exposito_Abur_Jaen2011}}. 
When system imbalance occurs, the zero and negative sequences are nonzero,
 and the
PMU’s output exhibits nonstationary frequency deviations \cite{Jouanne_Banerjee_2001}, \cite{Thorp444}.  In addition, the positive-sequence measurements become non-circular as described in \cite{Mandic_Sig_Proc_Mag}, \cite{Mandic}.
In the pioneering works of \cite{Mandic_Sig_Proc_Mag} and \cite{Mandic},  new methods were derived for frequency-estimation based on  non-circular models and the Clarke’s transformation. These methods   use  the positive and negative sequences and  analyze the measurements   in the time domain.
 The  mismatch estimation error caused by using the balanced state estimation  under imbalance
is studied in
\cite{Abur2002} and the influence of imperfect synchronization on the state estimation is described in 
\cite{Imperfect_Sync}.
In \cite{Monti_2013}
a distribution system state estimator  suitable for  monitoring unbalanced distribution networks is presented. 
A practical procedure to decrease the state estimation error introduced by load imbalances is developed in
\cite{Tuykom_Maun_Abur2010}.

In the literature,  various definitions are given for imbalance in a  power system, where the  fundamental performance measures are the  voltage unbalance factor (VUF) \cite{Jouanne_Banerjee_2001}, \cite{unbalanced_def}, \cite{unbalanced_defs} and the percent voltage unbalance (PVU) 
\cite{Singh_Singh_Mitra2007}.
 The VUF is the ratio of the magnitudes of negative- and positive-sequence voltages and the PUV is equal to the ratio of the maximum voltage magnitude deviation  of the zero, positive, and negative sequences from the average of the three-phase voltage magnitudes  \cite{Lee1999}.
The phase angle imbalance, which is not reflected in either the VUF or PVU measures,
can be  described by 
the phase voltage unbalance factor (PVUF) \cite{1575107} and the complex VUF (CVUF) \cite{Pillay_Hofmann2001}, \cite{Faiz_Ebrahimpour_Pillay2004}. 
The limitations of these commonly-used methods  can be found, for example, in \cite{Filho_2010}. 
An	online identification method of the level, location, and effects of voltage
imbalance in a  distribution network is derived in \cite{Woolley_Milanovic_2012}
based on distribution system state estimation.
 However, the existing non-parametric methods for detection of imbalance  are insufficient
(e.g., {\cite{Filho_2010,Pillay_Hofmann2001,Faiz_Ebrahimpour_Pillay2004,Faiz_Ebrahimpour2005}}). Derivation of parametric detection methods is expected to improve the detection performance.

A particularly relevant prior work is \cite{detection2013} where the authors develop the first parametric GLRT for detecting voltage and phase imbalances based on time domain measurements.
  While \cite{detection2013} and this paper both use GLRT principle, the models considered and the statistics used are quite different. 
	Specifically, the approach presented here is based on the native PMU (frequency domain) output that is less informative than the time domain measurements used in \cite{detection2013} but more readily accessible\footnote{For example, the proposed method is able to detect imbalances based on $K=2$ frequency domain samples. These samples are based on compressed $N+1\gg K$ time domain samples.}.  More significance, perhaps, is the formulation of hypotheses. 
	The GLRT derived in \cite{detection2013} tests the hypothesis that the system is {\em{perfectly}} balanced against any amount of imbalance in the system. Our approach, on the other hand, aims to detect substantial imbalance.  The presence of imbalance in the null hypothesis (the nominal case) presents nontrivial technical difficulties, which cannot be dealt with by simply changing the detection threshold on a test  designed for a perfectly balanced system operating at the nominal frequency.
	Finally, since usually the zero-sequence power does not propagate to the machine terminals
	\cite{Jouanne_Banerjee_2001}, {\cite{Arrillaga_Bollen_Watson_2000,4636745,Davoudi_2012}},
	the information which is used by the proposed GLRT 
includes only  the  positive and negative sequence components, while \cite{detection2013} uses the three sequences and investigates  the influence of the zero sequence on the detection performance.

In most situations,  frequency deviations and {\em{minor}} imbalances can be mitigated by frequency regulation 
	or load compensation techniques \cite{Ghosh_Joshi}. 
In the literature, several mitigation techniques have been suggested to correct {\em{significant}} voltage imbalance problems \cite{Jouanne_Banerjee_2001}, 
on both the  power system and user facility levels.
Voltage imbalance is ultimately fixed by manually or automatically rebalancing loads and removing asymmetric network line configurations \cite{Woolley_Milanovic_2012}, where these are costly processes and inappropriate for frequent but small imbalances.
For example,   the compensation of the voltage imbalance can be achieved by reducing the negative-sequence voltage using a series active power filter or based on shunt compensation, as described in 
\cite{Savaghebi_Jalilian_Vasquez_Guerrero2013}, 
or by advanced control strategies  {\cite{Bidram_Davoudi_2012,Salama_Elnady_2005,Meersman2011899}}.  In addition, the compensation of voltage harmonics, which can be generated by a nonlinear unbalanced load, can be considered by separating the positive and negative sequences of each harmonic order \cite{Savaghebi_Jalilian_Vasquez_Guerrero2013}. 
\begin{table}
{\small{\em{Nomenclature}}}\\
\\
\begin{tabular}{p{2.65cm} p{5.5cm}}
  $V_{a}, V_{b},V_{c}$ & Three-phase   voltage magnitudes  \\
  $\varphi_a,\varphi_b,\varphi_c$ & Three-phase   voltage phases  \\
	$\vvec$ & $\left[V_a e^{j \varphi_a},V_b e^{j \varphi_b},V_c e^{j \varphi_c}
\right]^{\mbox{\tiny T}}$
\\
  $\omega_0$ & Nominal grid-frequency   \\
	$\Delta$& Frequency deviation\\
	$\hat{\Delta}^{(i)}$& ML frequency-deviation estimator under hypothesis $i=0,1$\\
	$\hat{\Delta}_{s}$& Suboptimal frequency-deviation estimator
\\
	$N$ & Samples  per cycle at time domain\\
$K$  &  Samples  at the frequency domain \\
$v_a[n],v_b[n],v_c[n]$& Three-phase voltages   at time $n$ \\
$V_0[k],V_+[k],V_-[k]$ &   Frequency domain zero, positive, and negative  phasor sequences   at time $k$
\\
$\wvec_{a,b,c}[n]$& Gaussian noise sequences at time $n$ \\
$W_0[k],W_+[k],W_-[k]$&
  Complex  Gaussian noise sequences at frequency-time $k$\\
 $\sigma^2$ & Noise variance
\\
$\Imat_K$ & $K \times K$ identity matrix\\
${\mathbf{1}}_K $& Vector of ones of size $K$\\
 $\alpha$ & $e^{j 2\pi/3}$ \\
$\gamma$&  $\frac{2\pi}{N}$\\
$C_0,C_+,C_-$ & Zero, positive, and negative phasors\\
$\hat{C}_+^{(i)},\hat{C}_-^{(i)}$ &Positive and negative phasor CML estimators under hypothesis $i$\\
$\hat{C}_-^{(uc)}$ &  Negative phasor ML estimator 
\\
$\tilde{\bnu}_0,\tilde{\bnu}_+,\tilde{\bnu}_-$ &  PMU's measurement vectors 
\\
$\bnu_0,\bnu_+,\bnu_-$ & Whitened PMU's measurement vectors \\
$\thetavec$ & Unknown parameters vector\\
$L(\thetavec)$ & Likelihood function at $\thetavec$ \\ 
$T_{\text{GLRT}}$&  GLRT detector\\
$T_{\text{GLRT-SNH}}$&  GLRT-SNH detector \\
$T_{\text{VUF}}$ &  VUF detector \\
 $\tau$ &   Detector's threshold\\
$P_e(\tau)$ & GLRT false alarm probability at  $\tau$ \\
$P_e^{(a)}(\tau)$& False alarm asymptotic probability at  $\tau$\\
$r$& Authorized level of imbalances.
\end{tabular}
\end{table}
\subsection{Organization and notations}
 The remainder of the paper is organized as follows:
Section \ref{model} presents the mathematical model  and outlines several special cases.
The GLRT  detector and CML estimators for detecting imbalance  are derived in Section \ref{GLRT_sec}.
In Section \ref{theoretical_sec}, a performance analysis of the proposed GLRT is developed.
Finally, the proposed method is evaluated via simulations in Section \ref{simulation_sec}
and the conclusion appear in Section \ref{diss}.

In the rest of this paper we denote vectors  by boldface
lowercase letters and matrices  by boldface uppercase
letters.
The operators $(\cdot)^*$,
 $(\cdot)^{\mbox{\tiny $T$}}$, $(\cdot)^{\mbox{\tiny $H$}}$, and   $(\cdot)^{-1}$ denote the complex conjugate, transpose, Hermite, and inverse operators,  respectively.
The operator ${\text{Real}}\{\cdot\}$  denotes the real  part of its argument.
 For
convenience, variables are cataloged in the Nomenclature Table.
\section{Measurement model}
\label{model}
The system and measurement models considered here are 
conventional (see, e.g.,  \cite{PMUbook,Thorp444}).
In this section we present  the model in a statistical signal processing formulation
that includes a  description of the noise statistics, and it is more
 convenient for developing estimation and detection algorithms \cite{ICASSP_unbalanced}.  
In particular, we  describe the  statistical behavior of
the PMU output,  i.e. after the sampling, symmetrical  transformation,
and  nominal-frequency discrete Fourier transform (DFT)  operation.
\subsection{Off-nominal  unbalanced system phasors}
The    voltages in a three-phase  power system are  assumed to be pure sinusoidal signals
  of frequency $\omega_0+\Delta$, 
  where
  $\omega_0$
is the known nominal-frequency ($100\pi$ or $120\pi$) and 
$\Delta$ is the frequency deviation from this nominal value.
The magnitudes and phases  of the three   voltages are denoted by
$V_{a}, V_{b},V_{c}\geq 0$ and $\varphi_a,\varphi_b,\varphi_c\in[0,2\pi]$, respectively.
The three-phase power system is  balanced or symmetrical  if $V_a=V_b=V_c$ and 
$\varphi_a=\varphi_b+ \frac{2\pi}{3}=\varphi_c-\frac{2\pi}{3}$.
The PMU    samples these real signals  $N$  times per cycle of 
the nominal-frequency, $\omega_0$,  to produce the following 
 discrete-time, noisy measurements  model  (e.g. \cite{PMUbook}, pp. 51-52 and 
\cite{Phasor}):
\beqna
\label{phasor_model2_three}
\left[\begin{array}{c}v_a[n]\\v_b[n]\\v_c[n]
\end{array}
\right]
=\left[ \begin{array}{c} V_a\cos\left(
\gamma \frac{\omega_0+\Delta }{\omega_0}n+\varphi_a\right)\\
V_b\cos\left(
\gamma \frac{\omega_0+\Delta }{\omega_0}n+\varphi_b\right)\\
V_c\cos\left(
\gamma \frac{\omega_0+\Delta }{\omega_0}n+\varphi_c\right)\end{array}
\right]+\wvec_{a,b,c}[n]\hspace{0.1cm}
\nonumber\\
={\text{Real}}\left\{ e^{
j\gamma \frac{\omega_0+\Delta }{\omega_0}n}\vvec\right\}+\wvec_{a,b,c}[n]\hspace{1.9cm}
\nonumber\\
=
\frac{1}{2}e^{
j\gamma \frac{\omega_0+\Delta }{\omega_0}n}\vvec
+\frac{1}{2}e^{-j\gamma \frac{\omega_0+\Delta }{\omega_0}n}\vvec^*
+\wvec_{a,b,c}[n]
,
\eeqna
 for all $ n\in{\mathbb{Z}}$,
where 
 $\gamma\define \frac{2\pi}{N}$ and  
$\vvec\define \left[V_a e^{j \varphi_a},V_b e^{j \varphi_b},V_c e^{j \varphi_c}
\right]^{\mbox{\tiny T}}$.
The noise sequence,  
$\{\wvec_{a,b,c}[n]\}_{n\in{\mathbb{R}}}$, is assumed to be a
real  white Gaussian noise sequences with 
known covariance matrix  $\sigma^2 \Imat_3$.
The derived method can be easily extended to the more general case of a
 correlated three-phase system \cite{Woolley_Milanovic_2012} by using a non-diagonal covariance matrix.
The error covariance matrix can be obtained, for example, as described in \cite{Woolley_Milanovic_2012}.

The PMU constructs  the  complex representation 
 of  the  signals
by using a DFT operator 
 over one cycle of the
nominal-frequency
 \cite{PMUbook}, \cite{Thorp444}.
 That is, the PMU DFT operation on any arbitrary signal $x[n]$ results in the following phasor sequence:
 \beqna
\label{phasor_model}
X[k] =\frac{\sqrt{2}}{N}\sum_{n=k}^{k+N-1}x[n]e^{-j \gamma n},~~~k=0,\ldots, K-1,
\eeqna
where the index $k$ refers to the beginning of the DFT window.
By substituting  the three  sequences, $v_a[n]$,
$v_b[n]$, and
$v_c[n]$, for all $n\in{\mathbb{Z}}$  from (\ref{phasor_model2_three}) in (\ref{phasor_model}).
Using
 the identity \cite{framework},
\[
\sum_{n=k}^{k+N-1} e^{j  \alpha n}
=\frac{\sin (\alpha N/2)}{\sin (\alpha /2)}
e^{j \alpha\left(k+\frac{N-1}{2}\right)},
~\forall \alpha\in{\mathbb{R}},
\]
  we obtain the following
  phasor sequences measurements:
	\beqna
\label{phasor_model2_three3}
\left[\begin{array}{c}V_a[k]\\V_b[k]\\V_c[k]
\end{array}
\right]&=&
\frac{\sqrt{2}}{N}\left[\begin{array}{c}\sum_{n=k}^{k+N-1}v_a[n]e^{-j \gamma n} \\\sum_{n=k}^{k+N-1}v_b[n]e^{-j \gamma n}\\\sum_{n=k}^{k+N-1}v_c[n]e^{-j \gamma n}
\end{array}
\right]
\nonumber\\
&=&
\frac{1}{\sqrt{2}}P
e^{j \gamma\frac{\Delta }{\omega_0}k}\vvec
+\frac{1}{\sqrt{2}}Q
 e^{-j \gamma\frac{2\omega_0+\Delta }{\omega_0}k}\vvec^*\nonumber\\&&
+\frac{\sqrt{2}}{N}\sum_{n=k}^{k+N-1}\wvec_{a,b,c}[n]e^{-j \gamma n}
,
\eeqna
$k=0,\ldots,K-1$,  where 
\beqna
\label{P}
P&=&\frac{\sin (\gamma N\frac{\Delta }{2\omega_0})}{N\sin (\gamma\frac{\Delta }{2\omega_0})}
e^{j \gamma\frac{\Delta }{\omega_0}\frac{N-1}{2}}
\\
\label{Q}
Q&=& \frac{\sin (\gamma N\frac{2\omega_0+\Delta }{2\omega_0})}{N\sin (\gamma\frac{2\omega_0+\Delta }{2\omega_0})}
e^{-j \gamma\frac{2\omega_0+\Delta }{\omega_0}\frac{N-1}{2}}.
\eeqna
It is seen, then,  that $P$ and $Q$ are functions of the unknown frequency deviation,  $\Delta$,  but independent of $k$.

Finally,
the three symmetrical voltage sequences    are calculated  from three-phase voltages by the PMU using
the symmetrical  component transformation  (e.g. \cite{PMUbook} pp. 63-67):
\be
\label{positive_seq}
\left[\begin{array}{c}V_0[k]\\
V_+[k]\\
V_-[k]
\end{array}
\right]=\frac{1}{3}
\underbrace{
\left[\begin{array}{ccc}1&1&1\\
1&\alpha&\alpha^2\\
1&\alpha^2&\alpha
\end{array}
\right]}_{\define 
\Hmat}\left[\begin{array}{c}V_a[k]\\
V_b[k]\\
V_c[k]
\end{array}
\right]
,
\ee
for all $ k=0,1,\ldots,K-1$,
where 
 $V_0[k],
V_+[k]$, and
$V_-[k]$ are the zero, positive, and negative sequences, respectively,  and $\alpha=e^{j 2\pi/3}$.
By substituting (\ref{phasor_model2_three3}) in (\ref{positive_seq}), we obtain
	\beqna
\label{phasor_model2_zero}
V_0[k]=P
e^{j \gamma\frac{\Delta }{\omega_0}k}
C_0 +Q
 e^{-j \gamma\frac{2\omega_0+\Delta }{\omega_0}k}
C_0^*+W_0[k]
\eeqna
	\beqna
\label{phasor_model2}
V_+[k]=P
e^{j \gamma\frac{\Delta }{\omega_0}k}
C_+ +Q
 e^{-j \gamma\frac{2\omega_0+\Delta }{\omega_0}k}
C_-^*+W_+[k]
\eeqna
	\beqna
\label{phasor_model2_neg}
V_-[k]=P
e^{j \gamma\frac{\Delta }{\omega_0}k}
C_- +Q
 e^{-j \gamma\frac{2\omega_0+\Delta }{\omega_0}k}
C_+^*+W_-[k]
\eeqna
$k=0,\ldots,K-1$,
where
\[ \left[C_0,
C_+,
C_-
\right]^T
=\frac{\sqrt{2}}{6}\Hmat
\vvec
\]
and
\[ \left[W_0[k],
W_+[k],
W_-[k]
\right]^{\mbox{\tiny $T$}}
=\frac{\sqrt{2}}{3N}\Hmat
\sum_{n=k}^{k+N-1}\wvec_{a,b,c}[n]e^{-j \gamma n},
\]
for all $ k=0,\ldots,K-1$.
Since  $\Hmat\Hmat^{\mbox{\tiny $H$}}=3\Imat_3$, 
the  noise sequences, $W_0[k]$, $W_+[k]$, $W_-[k]$, $k=1,\ldots, K-1$,
 are  independent 
complex circularly symmetric   Gaussian noise sequences where each sequence has a variance of   $\frac{ 2 \sigma^2}{3 N } $.
However, it should be noted that if the original three-phase noise signals are correlated,
which is the case in  distribution systems \cite{Woolley_Milanovic_2012}, \cite{wang_2001},
 then, the
noise sequences of the three symmetrical components are also correlated.
It can be seen that the PMU output in (\ref{phasor_model2_zero})-(\ref{phasor_model2_neg})
 includes  samples  of the  symmetrical sequences, $W_0[k],W_+[k],W_-[k]$, at  the nominal-frequency bin, that are different
from the true value of the input  sequence phasors, $C_0$, $C_+$, and $C_-$.

In this work, 
we are interested in the detection of
imbalances
 based  on $K$ measurements of the positive and negative sequences
 from
(\ref{phasor_model2_zero})-(\ref{phasor_model2_neg}).
The PMU output of the zero sequence, $V_0[0],\ldots,V_0[K-1]$
is  usually non observable and is described in this paper for the sake of completeness.
 The models for these $K$ measurements  can be written in  matrix form as follows:
 \beqna
\label{phasor_model234}
\tilde{\bnu}_0 &=&P
C_0 \tilde{\evec}_1 
+Q C_0^* \tilde{\evec}_2
+\tilde{\wvec}_0\\
\label{phasor_model234+}
\tilde{\bnu}_+ &=&P
C_+ \tilde{\evec}_1
+Q C_-^* \tilde{\evec}_2
+\tilde{\wvec}_+\\
\label{20}
\tilde{\bnu}_- &=&
P C_- \tilde{\evec}_1+Q
C_+^* \tilde{\evec}_2 
+\tilde{\wvec}_-,
\eeqna
where 
\beqna
\tilde{\bnu}_0&\define&[V_0[0],\ldots,V_0[K-1]]^T,
\nonumber\\
\tilde{\bnu}_+&\define&[V_+[0],\ldots,V_+[K-1]]^T, 
\nonumber\\
 \tilde{\bnu}_-&\define&[{V}_-[0],\ldots,{V}_-[K-1]]^T,\nonumber
\\
\tilde{\evec}_1&\define&\left[1,e^{j \gamma\frac{\Delta }{\omega_0}},\ldots,e^{j \gamma\frac{\Delta }{\omega_0}(K-1)}\right]^T,
\nonumber\\
\tilde{\evec}_2&\define&\left[1, e^{-j \gamma\frac{2\omega_0+\Delta }{\omega_0}},\ldots, e^{-j \gamma\frac{2\omega_0+\Delta }{\omega_0}(K-1)}\right]^T.
\nonumber
\eeqna
The vectors in $\tilde{\evec}_1$ and $\tilde{\evec}_2$ are identical to 
the steering vector for a uniform  linear array \cite{Kay_detection}.
The noise vectors,
 $\tilde{\wvec}_0$,
,$\tilde{\wvec}_+$,
 and $\tilde{\wvec}_-$, are independent zero-mean
complex, circularly symmetric, {\em{colored}}    Gaussian noise sequences with
 covariance matrix $\Rmat $, where $\Rmat$ is a
 $K\times K $ matrix with 
the following $(k,l)$th element 
 \[[\Rmat]_{k,l} =\frac{2\sigma^2}{3N^2}\left\{\begin{array}{lr}
  N-|k-l| & {\text{if }}-N\leq k-l\leq N\\
  0& {\text{otherwise}}
  \end{array}\right. .
\]
Since the error covariance matrix is known, the signals in (\ref{phasor_model234})-(\ref{20}) can be
 prewhitened. The whitening
operation is performed by left-multiplication of the terms in (\ref{phasor_model234})-(\ref{20}) by
$\Rmat^{-\frac{1}{2}}$:
\beqna
\label{phasor_model234w}
\bnu_0 &=& \Rmat^{-\frac{1}{2}}
\tilde{\bnu}_0=P
C_0{\evec}_1
+Q C_0^* {\evec}_2
+{\wvec}_0
\\
\label{only_positive}
\bnu_+ &=&\Rmat^{-\frac{1}{2}}
\tilde{\bnu}_+=P
C_+ {\evec}_1
+Q C_-^* {\evec}_2
+{\wvec}_+
\\
\label{20w}
\bnu_- &=&\Rmat^{-\frac{1}{2}}
\tilde{\bnu}_-=
P C_-{\evec}_1+Q
C_+^* {\evec}_2 
+{\wvec}_-,
\eeqna
where ${\evec_m}\define \Rmat^{-\frac{1}{2}}\tilde{\evec}_m  $, $m=1,2$.
The modified noise vectors, ${\wvec}_0= \Rmat^{-\frac{1}{2}}
\tilde{\wvec}_0 $, ${\wvec}_+= \Rmat^{-\frac{1}{2}}
\tilde{\wvec}_+$, ${\wvec}_-= \Rmat^{-\frac{1}{2}}
\tilde{\wvec}_- $ have 
an identity covariance matrix, $\Imat_K$.
Similarly, the prewhitening procedure  can be performed for the more general case of a
 correlated three-phase system, when  the three sequences are dependent.

\subsection{Special cases}
\subsubsection{Perfectly balanced system}
For the special case of a {\em{perfectly}} balanced system, the three-phase voltages satisfy
$
V_a =V_b=V_c $ and $\varphi_a=\varphi_b+ \frac{2\pi}{3}=\varphi_c-\frac{2\pi}{3}$.
Therefore, it can be verified that for this case
 $C_0=0$,  $C_-=0$, and
the model in (\ref{phasor_model234w})-(\ref{20w}) is reduced to
\beqna
\label{phasor_model234b}
\left\{\begin{array}{l}
\bnu_0 =\wvec_0\\
\bnu_+ =P
C_+ \evec_1 
+\wvec_+\\
\bnu_- =
Q C_+^* \evec_2
+\wvec_-
\end{array}\right. .
\eeqna
The model in (\ref{phasor_model234b}) indicates that
for  perfectly balanced systems the zero-sequence is a noise-sequence and
the positive and negative sequences create sinusoidal signals.

\subsubsection{Nominal-frequency system}
\label{nominal_case}
If the input signal is a  pure sinusoid at the nominal-frequency, i.e. $\Delta=0$, then,
  by using  (\ref{P}),  (\ref{Q}), and
	the L'H$\hat{\text{o}}$pital's rule, it can be seen that 
  $\lim\limits_{\Delta\rightarrow 0}P=1$, $\lim\limits_{\Delta\rightarrow 0}Q=0$, and $\tilde{\evec}_1  ={\mathbf{1}}_K$.
By substituting these terms in (\ref{phasor_model234w})-(\ref{20w}), the output of the PMU
  for the nominal-frequency case is given by 
   \beqna
   \label{positive_seq_comp_nom}
   \left\{
\begin{array}{l}
\bnu_0 =
C_0 \Rmat^{-\frac{1}{2}}{\mathbf{1}}_K
+\wvec_0\\
\bnu_+ =
C_+ \Rmat^{-\frac{1}{2}}{\mathbf{1}}_K
+\wvec_+\\
\bnu_- =C_- \Rmat^{-\frac{1}{2}}{\mathbf{1}}_K
+\wvec_-
\end{array}\right.
.
\eeqna
For a perfectly balanced system at the nominal-frequency 
we substitute  $C_-=C_0=0$ in (\ref{positive_seq_comp_nom}) and obtain
 \beqna
\label{positive_seq_comp_nom2}
   \left\{
\begin{array}{l}
\bnu_0 =\wvec_0\\
\bnu_+ =
C_+ \Rmat^{-\frac{1}{2}} {\mathbf{1}}_1
+\wvec_+\\
\bnu_- =\wvec_-
\end{array}\right.
.
\eeqna
Therefore, it can be seen from (\ref{positive_seq_comp_nom}) and (\ref{positive_seq_comp_nom2})  
that for a system operated at nominal-frequency, system imbalance is undetectable 
using only the  positive-sequence phasors
since the  model of the positive-sequence, $\bnu_+$, is identical under both circumstances.
This is in contrast to  state estimation
and signal analysis, which are carried out using
only the positive-sequence model {\cite{Abur2002,Exposito_Abur_Jaen2011}}.
Thus, in order to detect unbalanced situations versus perfectly unbalanced situation we should also use  the zero- and/or the negative-sequence\footnote{It should be noted that at off-nominal frequency a detection method can be derived  based only on the model of the positive-sequence in (\ref{only_positive}), similar to the derivations of the GLRT in Section \ref{GLRT_sec}. However, since 
typical magnitudes of $Q$ are small compared to $P$ (Chapter 3 in \cite{PMUbook}), imbalance is detectable   solely by using $\bnu_+$  only for significantly high 1)  signal-to-noise-ratio (SNR); and/or 2) number of samples; and/or 3) frequency deviations.
}.

\section{Detection of imbalance and the GLRT}
\label{GLRT_sec}
\label{detection_section}
\subsection{The hypothesis-testing problem}
The objective of this study is to develop a  method for
significant system imbalance detection based on the PMU output.
In most phasor measurement applications, it is common to have some degree of imbalance in the system
 \cite{ANSI_C84.1}. 
Therefore it is important that the detector is robust to  modest level of imbalance.
Furthermore, in the vast majority of cases, the zero-sequence signal does not propagate to the machine terminals \cite{Jouanne_Banerjee_2001},
{\cite{Arrillaga_Bollen_Watson_2000,4636745,Davoudi_2012}}. Therefore 
the problem of imbalance detection is developed in this section based only on the whitened positive and negative sequence components.
For the special case of independent three-phase signals,  the three symmetrical measurements sequences in (\ref{phasor_model234w})-(\ref{20w})  are also  independent and 
the problem of imbalance  detection based on the positive and negative sequences is independent of the zero-sequence.

The  detection problem can be formulated  as the following composite hypothesis testing problem:
\begin{equation}
\label{detection_prob}\left\{
{\begin{array}{l}
H_0:  |C_-|^2\leq   r^2
\\
H_1:  |C_-|^2> r^2  
\end{array}}\right. 
\end{equation}
where  $r$ is an authorized level of imbalance,
and  hypotheses $H_0$ and $H_1$ represent the balanced and imbalanced hypothesis, respectively. 
That is, the measurement model under either hypotheses is given in (\ref{phasor_model234w})-(\ref{20w}),
i.e., the   likelihood functions are identical, and the difference between the hypotheses is  the magnitude of $|C_-|$.
This problem is known in the literature as a constrained hypothesis testing  problem \cite{Moore}.

The detection problem in (\ref{detection_prob}) is a composite test, i.e.  the measurement likelihood functions depend on unknown parameters, $C_+$, $C_-$,
and 	 $\Delta$.
Hence,   the GLRT is a natural choice for this problem.
The GLRT adopts the general alternative $H_1$ against
$H_0$ if the ratio of the likelihood functions is greater than a threshold,
where the unknown parameters are replaced by their respective maximum-likelihood (ML) estimators
 \cite{Kay_detection}.
In the presence of parametric constraints, 
the ML estimators should be replaced by 
the CML estimators  \cite{Moore}.

 \subsection{State estimation}
Let $\hat{\thetavec}_i^{(i)}$  denote the  ML estimator of $\thetavec
=[C_{+}, C_-, \Delta]^{\mbox{\tiny T}}$ under hypothesis $i$ 
and 
 $f(\bnu_+,\bnu_-;\thetavec)$ is the   probability density function (pdf) of $\bnu_+$ and $\bnu_-$. 
Based on the model described in (\ref{phasor_model234w})-(\ref{20w}),
the likelihood function is given by
 \beqna
\label{fH1}
L(\thetavec)&\define& \log f(\bnu_+,\bnu_-;\thetavec)\nonumber\\&=&2 K\log \pi-
\left|\left|
{\bnu}_+ -P 
C_+ {\evec}_1 
-Q C_-^* {\evec}_2\right|\right|^2
\nonumber\\
&&-\left|\left|{\bnu}_- -  
 P C_- {\evec}_1-
QC_+^* {\evec}_2 
\right|\right|^2.
\eeqna
In this Section, we develop the CML estimators under the balanced/unbalanced system constraints.
\subsubsection{The CML  estimators for balanced systems}
Under the balanced system  constraint,
$|C_-|^2 \leq   r^2$,
the CML estimator of  $\thetavec$ under $H_0$
is given by
\beqna
\hat{\thetavec}^{(0)}=\arg\max_{\thetavecsmall}L(\thetavec)~~~
\mbox{subject to}~~~|C_-|^2 \leq   r^2\nonumber.
\eeqna
Therefore, under $H_0$
we maximize the   following
Lagrangian:
\beqna
\label{Q1_lag_new}
Q_0=L(\thetavec)-\mu_{0}^2\left( |C_-|^2-r^2\right),
\eeqna
where   $\mu_{0}^2$  is the 
  Karush-Kuhn-Tucker (KKT)  multiplier \cite{Boyd_2004} under $H_0$.
For a fixed $\Delta$, 
 by equating the complex derivatives  of the  right hand side (r.h.s.) of (\ref{Q1_lag_new})
with respect to (w.r.t.)  $C_+$ and $C_-$
 to zero, one obtains
  \beqna
    \label{Cp_hatH0}
 \hat{C}_+^{(0)} &=&\frac{z_+-\kappa_2 (\hat{C}_-^{(0)})^* 
}{\kappa_1}
\\
    \label{Cn_hatH0}
\hat{C}_-^{(0)}&=&\frac{z_-
- \kappa_2(\hat{C}_+^{(0)})^*
}
{\kappa_1+\mu_{0}^2},
\eeqna
where
$
z_+\define
P^*{\evec}_1^H{\bnu}_+
+
Q{\bnu}_-^H \evec_2,
$
$z_-\define
P^*\evec_1^H{\bnu}_-+Q{\bnu}_+^H\evec_2,
$
$
\kappa_1\define {|P|^2\evec_1^H \evec_1 +|Q|^2\evec_2^H{\evec}_2
 },
$
and
$\kappa_2\define 2P^*Q  {\evec}_1^H{\evec}_2 .
$
By using some mathematical manipulations, the CML estimators  in (\ref{Cp_hatH0})-(\ref{Cn_hatH0}) can be rewritten as:
\beqna
\label{Cp_hat_fH0}
 \hat{C}_+^{(0)} &=&\frac{(\kappa_1+\mu_0^2)z_+-\kappa_2 z_-^*}{(\kappa_1+\mu_0^2)\kappa_1-|\kappa_2|^2} 
\\
\label{Cn_hat_fH0}
 \hat{C}_-^{(0)} &=&
\frac{\kappa_1z_--\kappa_2z_+^* }{(\kappa_1+\mu_0^2)\kappa_1-|\kappa_2|^2}.  
\eeqna
It should be noted that, according to (\ref{Cp_hat_fH0}) and (\ref{Cn_hat_fH0}),
 the {\em{magnitude}} constraints  have no influence on the {\em{phase}} of the estimator
  $
 \hat{C}_-^{(0)}$.
By using the  primal feasibility and complementary slackness KKT conditions \cite{Boyd_2004}, it can be shown that the KKT multiplier satisfies:
\beqna
\label{mu_minus}
\mu_{0}^2=
\left\{\begin{array}{lr}0 &{\text{if }} \left|\hat{C}_-^{(uc)}\right|^2  \leq r^2
\\\frac{\kappa}{r}\left(\left|\hat{C}_-^{(uc)}\right|-r
\right)
&{\text{otherwise}}
\end{array}\right.,
\eeqna
where
\be
\label{uc_est}
\hat{C}_-^{(uc)}\define \frac{\kappa_1z_--\kappa_2z_+^*}{\kappa_1^2-|\kappa_2|^2},
\ee
which is the unconstrained ML estimator of the negative phasor,
and
$\kappa\define \frac{\kappa_1^2-|\kappa_2|^2}{\kappa_1}$.
By substituting  (\ref{mu_minus}) in 
(\ref{Cn_hat_fH0}),  one obtains
the  CML negative phasor estimator:
\beqna
\label{new_Cp_neg_fH0}
\hat{C}_-^{(0)} =\left\{\begin{array}{lr} \hat{C}_-^{(uc)} &{\text{if }} \left|\hat{C}_-^{(uc)}\right|^2  \leq r^2
\\r\frac{\hat{C}_-^{(uc)}}{\left|\hat{C}_-^{(uc)}\right|}
&{\text{otherwise}}
\end{array}\right.,
\eeqna
where the positive-sequence phasor can be calculated by substituting (\ref{new_Cp_neg_fH0})
in (\ref{Cp_hatH0}).

\subsubsection{The CML  estimators for unbalanced systems}
Similarly, under the imbalanced system hypothesis, $H_1$, 
the CML estimator of  $\thetavec$ 
is given by
\beqna
\label{Q2_lag_new}
\hat{\thetavec}^{(1)}=\arg\max_{\thetavecsmall}L(\thetavec)~~~
\mbox{subject to}~~~|C_-|^2 >   r^2\nonumber.
\eeqna
Similar to the derivations of the CML estimators for the balanced system in (\ref{Cp_hatH0}) 
and (\ref{new_Cp_neg_fH0}),  it can be shown that the solution of the maximization in (\ref{Q2_lag_new})  is given by
  \beqna
    \label{Cp_hatH1}
 \hat{C}_+^{(1)} &=&\frac{z_+-\kappa_2 (\hat{C}_-^{(1)})^* 
}{\kappa_1}
\\
\label{new_Cp_neg_fH1}
\hat{C}_-^{(1)} &=&\left\{\begin{array}{lr} \hat{C}_-^{(uc)} &{\text{if }} \left|\hat{C}_-^{(uc)}\right|^2  > r^2
\\r\frac{\hat{C}_-^{(uc)}}{\left|\hat{C}_-^{(uc)}\right|}
&{\text{otherwise}}
\end{array}\right..
\eeqna

\subsubsection{Frequency-deviation estimation}
If the frequency deviation is unknown then, the phasor estimators
$\hat{C}_+^{(i)}$ and  $\hat{C}_-^{(i)}$, $i=0,1$ are function of 
the frequency-deviation estimate. Similar to \cite{ICASSP_unbalanced}, it can be shown that
the  ML frequency-deviation estimator is found 
by maximizing the likelihood function in (\ref{fH1}) after substituting the phasor CML estimators,
which results in
the following frequency-deviation ML estimator under $H_i$:
\beqna
\label{nice1s}
\hat{\Delta}^{(i)}=\arg\max_{\Delta}\left[\frac{|z_+|^2}{\kappa_1}-
\kappa \left|\hat{C}_-^{(i)}\right|^2
\right.
\nonumber\\
\left.
+2\kappa{\text{Real}}\left\{
\hat{C}_-^{(i)}(\hat{C}_-^{(uc)} )^*  \right\}
\right]
,
\eeqna
$i=0,1$.
 However, in practice, since the ML  frequency-deviation estimator  in (\ref{nice1s}) is based on a high complexity search, many other low-complexity frequency estimation methods are used in power systems (e.g. 
{\cite{PMUbook,Mandic_Sig_Proc_Mag,ICASSP_unbalanced,311162}}). In this work we use
the  state-of-the-art  frequency-estimation method, which  is based on the positive-sequence and given by \cite{Thorp444}:
\be
\label{method1}
\hat{\Delta}_{s}=\frac{\omega_0}{\gamma}\frac{1}{K}\sum\nolimits_{k=0}^{K-2}{\text{angle}}\left(V_+[k+1]\right)-{\text{angle}}\left(V_+[k]\right).
\ee

\subsection{The GLRT}
The  GLRT for the hypothesis-testing problem in (\ref{detection_prob}) 
declares
$H_1$ only if
$
T_{\text{GLRT}}$ is higher than a
given threshold, where the GLRT is given by \cite{Kay_detection}
\beqna
\label{GLRT_def}
T_{\text{GLRT}}\define 
L(\hat{\thetavec}^{(1)})-L(\hat{\thetavec}^{(0)})\hspace{1.8cm}
\nonumber\\
\label{GLRT_pn_org000}
=
L\left(\hat{\thetavec}^{(1)}=[\hat{C}_+^{(1)},\hat{C}_-^{(1)},\hat{\Delta}_{s}]^T\right)\hspace{0.1cm}
\nonumber\\
-L\left(\hat{\thetavec}^{(0)}=[\hat{C}_+^{(0)},\hat{C}_-^{(0)},\hat{\Delta}_{s}]^T\right),
\eeqna
where the last equality is under the assumption 
 that 
		we can
use the low-complexity frequency-deviation estimator, $\hat{\Delta}_{s}$, under both hypotheses.
By substituting  (\ref{fH1}) in (\ref{GLRT_pn_org000}), the GLRT   is given by
\beqna
\label{GLRT_pn_org}
T_{\text{GLRT}}=
-
\kappa_1\left(|\hat{C}_+^{(1)}|^2+|\hat{C}_-^{(1)}|^2-|\hat{C}_+^{(0)}|^2-|\hat{C}_-^{(0)}|^2\right)\hspace{0.6cm}
\nonumber\\+
2{\text{Real}}\left\{ \left(
\hat{C}_+^{(1)}   -
\hat{C}_+^{(0)}  \right)z_+^* +\left(
\hat{C}_-^{(1)}   -
\hat{C}_-^{(0)}  \right)z_-^*  \right.
\nonumber\\-\left.  \kappa_2^*\left(
\hat{C}_+^{(1)}\hat{C}_-^{(1)}   -
\hat{C}_+^{(0)}\hat{C}_-^{(0)}  \right)  \right\}.\hspace{2.5cm}
\eeqna
Substitution of  (\ref{Cp_hatH0}) and (\ref{Cp_hatH1}) in (\ref{GLRT_pn_org}), results in 
the following  GLRT detector:
\beqna
\label{GLRT_pn1}
T_{\text{GLRT}}=
\kappa \left(-\left|\hat{C}_-^{(1)}\right|^2
+\left|\hat{C}_-^{(0)}\right|^2\right.\hspace{2.8cm}
\nonumber\\
\left.
+2{\text{Real}}\left\{\left(
\hat{C}_-^{(1)}   -
\hat{C}_-^{(0)}  \right)\frac{\kappa_1z_-^*-\kappa_2^*z_+}{\kappa_1^2-|\kappa_2|^2}   \right\}\right).
\eeqna
By using (\ref{new_Cp_neg_fH0}) and (\ref{new_Cp_neg_fH1}),
the  GLRT can be rewritten as
\beqna
\label{GLRT_pn3}
T_{\text{GLRT}}=\kappa\left(|\hat{C}_-^{(uc)}|-r\right)^2
{\text{sign}}\left(|\hat{C}_-^{(uc)}|^2  - r^2\right)
,
\eeqna
where $\hat{C}_-^{(uc)}$ is defined in (\ref{uc_est}) and the ${\text{sign}}$ function is equal to $1$ for positive arguments and $-1$ otherwise.
Since the given threshold of the GLRT in (\ref{GLRT_def}) should be always nonnegative \cite{Kay_detection}, 
 the detector declares $H_0$  for any nonpositive $T_{\text{GLRT}}$.
Therefore,  the detector declares $H_0$ if $|\hat{C}_-^{(uc)}|^2  \leq r^2$. Thus,
 by applying a monotonically increasing transformation on the r.h.s. of
 (\ref{GLRT_pn3}),  the GLRT in this case 
decides $H_1$  if 
\beqna
\label{GLRT_pn}
T_{\text{GLRT}}=\sqrt{\kappa}\left(\left|\hat{C}_-^{(uc)}\right|-r\right)
>\tau,
\eeqna
where
$\tau\geq 0$.

The GLRT for detecting  imbalances in (\ref{GLRT_pn}) can be interpreted as a detector of
the presence 
of the  negative-sequence, which is consistent with the hypothesis testing as formulated in (\ref{detection_prob}).
That is, the detector  $T_{\text{GLRT}}$ is
proportional to  the unconstrained estimated negative phasor magnitude, while
the estimated phase of $\hat{C}_-^{(uc)}$  has no impact.
Since the positive-sequence appears in both the balanced and  unbalanced situations, 
the  positive-sequence phasor ML estimator is absent from the GLRT in
(\ref{GLRT_pn}).

\subsection{Special cases}
\subsubsection{Perfectly balanced system}
In the special case of a perfectly balanced system under $H_0$, i.e. 
 $r=0$,  
we obtained
the perfectly-balanced system CML estimators
$\hat{C}_-^{(0)} =0 $ and 
$\hat{C}_+^{(0)} =\frac{z_+
}{\kappa_1}$. By substituting  $r=0$  in (\ref{GLRT_pn}),
the  GLRT in this case decides $H_1$ only  if
\beqna
\label{GLRT_pn_perfect}
{T}_{\text{GLRT-SNH}}
=\sqrt{\kappa}\left|\hat{C}_-^{(uc)}\right|
>\tau.
\eeqna
The detector in (\ref{GLRT_pn_perfect}), named 
GLRT under simple null hypothesis (GLRT-SNH), detects a perfectly balanced versus unbalanced system.
Observing (\ref{GLRT_pn}) and (\ref{GLRT_pn_perfect}), 
it can be seen that the GLRT from (\ref{GLRT_pn_perfect}) 
can be rewritten as
\beqna
\label{connection}
T_{\text{GLRT}}=
{T}_{\text{GLRT-SNH}}
-\sqrt{\kappa}r.
\eeqna
Therefore, 
for a known frequency deviation, the detection of imbalances 
versus imperfect balanced system is identical to 
the detection of an imbalances 
versus a perfect balanced system with a shifted threshold by $\sqrt{\kappa}r$. As a result,
the  proposed GLRT can be enhanced by adjusting the threshold decision, taking
into account the desirable balance level,  $r$, and  the desirable
 probability of detection, which decreases as the threshold increases.
For estimated frequency deviation, however, $\sqrt{\kappa}r$ is a random parameter and these two detectors are different.

\subsubsection{Nominal-frequency system}
For $\Delta=0$, by using
  $P=1$ and $Q=0$, we obtain 
$z_-=\evec_1^H{\bnu}_-$,
$\kappa_1= \evec_1^H \evec_1$,
and
$\kappa_2=0$.
By substituting these values in (\ref{uc_est}), it can be verified that 
the unconstrained estimator of the negative phasor given for this case satisfies
\[
\hat{C}_-^{(uc)}=\frac{\evec_1^H{\bnu}_-}{\evec_1^H \evec_1}. 
\]
Thus, the estimator is only a function of the negative sequence. 
As a result,  the GLRT from (\ref{GLRT_pn}) in this case  is only a function of the negative sequence. 
This result is consistent with our result from Subsection \ref{nominal_case}:
For a system operated at nominal-frequency,  imbalance is undetectable when  based  on the  positive-sequence phasors, and the detection should be based  on the negative sequence.

\section{Theoretical performance analysis and threshold assessment}
\label{theoretical_sec}
In order to calibrate the test threshold  and analyze the detector's performance,
  the   GLRT distribution has to be determined. 
An exact analysis of the GLRT  in (\ref{GLRT_pn})
   is complicated
because of the nonlinear nature of the frequency-deviation estimator 
 and the  constrained  phasors   estimation.
 Therefore in this section we provide  a theoretical performance analysis of the GLRT  detector
for two special cases: 
1) known frequency deviation
 for the GLRT in (\ref{GLRT_pn});
 and
2)  asymptotic analysis.
for the GLRT-SNH in (\ref{GLRT_pn_perfect}).
These analyses  provide only an upper bound on the detection probability \cite{Kay_detection}.
\subsubsection{Performance analysis for known frequency deviation}
By using the independence between   $\bnu_+$ and $\bnu_-$ for a given frequency deviation, 
the unconstrained ML estimator of the   negative  phasor from
 (\ref{uc_est})  satisfies
\beqna
\label{pdf_C_m_uc}
\hat{C}_-^{(uc)}
\sim \left\{\begin{array}{lr} {\cal{CN}}\left(0,\frac{1}{\kappa}\right)&{\text{ Under }} H_0\\
{\cal{CN}}\left(C_-,\frac{1}{\kappa}\right) & {\text{ Under }} H_1
\end{array}\right.,
\eeqna
where  ${\cal{CN}}(\mu,\sigma^2)$  represents the complex circularly symmetric  Gaussian pdf with mean
$\mu\in{\mathbb{C}}$ and variance $\sigma^2\in{\mathbb{R}}$.
Therefore, 
the  GLRT-SNH  from (\ref{GLRT_pn_perfect})
admits  (e.g. \cite{Kay_detection} pp. 30-32):
\beqna
\label{chi_pm}
2{T}_{\text{GLRT-SNH}}
=
2 \sqrt{\kappa}\left|\hat{C}_-^{(uc)}\right|\hspace{3.25cm}\nonumber\\
\sim \left\{\begin{array}{lr} {\text{Rayleigh}}(1)&{\text{ Under }} H_0\\
{\text{Rician}}\left(4 {\kappa}\left|C_-\right|^2,1\right) & {\text{ Under }} H_1
\end{array}\right.,
\eeqna
where 
${\text{Rayleigh}}(\lambda)$
denotes the Rayleigh distribution with the mode parameter $\lambda$,
and ${\text{Rician}}(\left|C_-\right|^2,\lambda)$
denotes the  Rician distribution with the parameters  and $\lambda$.
By using (\ref{connection}) and (\ref{chi_pm}), the false alarm probability  for the  GLRT, i.e. 
the probability that  ${T}_{\text{GLRT}}$ is higher than a threshold
$\tau$ under $H_0$, is given by:
\beqna
\label{pe}
P_e(\tau)&=&\Pr\left(T_{\text{GLRT}}>\tau;H_0,\thetavec\right)
\nonumber\\
&=&\Pr\left(2T_{\text{GLRT-SNH}}
>2\tilde{\tau};H_0,\thetavec\right),
\eeqna
where, according to (\ref{connection}),
$\tilde{\tau}\define \tau+\sqrt{\kappa}r$.
By substituting the pdf from (\ref{chi_pm}) in (\ref{pe}), one obtains
\beqna
\label{GLRT_Pe}
P_e(\tau) =
\int_{2\tilde{\tau}}^\infty
x e^{-\frac{x^2}{2}}{\ud}x= e^{-\tilde{\tau }^2},~~~\forall \tau, r \geq 0. 
\eeqna
Inverting (\ref{GLRT_Pe}) gives the threshold for
the  GLRT detector,
where, by using this threshold,
the false alarm probability, $P_e(\tau)$, does not exceed
a predefined level.
The detection probability for the  GLRT detector,
$
\Pr\left({T}_{\text{GLRT}}>\tau;H_1,\thetavec\right)$, can be calculated in similar manner.

\subsubsection{Asymptotic performance for  the GLRT-SNH}
In this Subsection we consider the asymptotic (i.e., as $K$  tends to infinity) performance of the
 GLRT-SNH in (\ref{GLRT_pn_perfect}), $T_{\text{GLRT-SNH}}$.
In \cite{Kay_detection} pp. 205-206, it is shown that under suitable regularity conditions,
 the 
GLRT without any constraints, i.e. the GLRT-SNH with the ML frequency-deviation estimator in this case, 
 has the following probability of error
of the GLRT-SNH:
\beqna
\label{GLRT_Pe_a}
P_e^{(a)}(\tau) &\stackrel{a}{=}&
\Pr\left({T}_{\text{GLRT-SNH}}>\tau;H_0,\thetavec\right)
\nonumber\\
&=&
\int_{2{\tau}^2}^\infty \frac{1}{2} e^{-\frac{x}{2}}{\ud}x= e^{-\tau^2 }, 
\eeqna
for all $ \tau \geq 0$.
By comparing (\ref{GLRT_Pe}) and (\ref{GLRT_Pe_a}), it can be seen  that for large $K$ the 
GLRT-SNH performance
is the same whether  the frequency deviation is known or not.
Since the asymptotic pdf under $H_0$ does not depend on  the unknown parameters,
the threshold required to maintain a specific false alarm probability  can be found by (\ref{GLRT_Pe}).
This type of detector is referred to as a constant false alarm rate (CFAR) detector \cite{Kay_detection}.
However, the general GLRT detector, $T_{\text{GLRT}}$, is not CFAR since the threshold and the performance are also  functions of
$
\sqrt{\kappa}r$, which is function of 
the estimated frequency deviation.
In addition,
it should be noted that under unknown frequency deviation the noncentrality parameter
 is decreased, hence the reduction of detection probability. This can be interpreted as information reduction caused by the need to estimate additional parameters for use in the detector
 \cite{Kay_detection}.

 \section{Simulations}
 \label{simulation_sec}
 In this section,  the performances of the ML frequency deviation estimation and the  proposed GLRT in (\ref{GLRT_pn})
are evaluated.
We consider a single PMU and a
sampling rate of $N=48$ samples per cycle of the
nominal grid frequency, $\omega_0=2\pi \cdot 60$,
and  $K=12$  frequency samples. 
The performance is
evaluated using $5000$ Monte-Carlo simulations.
Unless otherwise specified,
the frequency of the input signal is assumed to have a $\Delta=0.1\times 2\pi $ offset from the 
nominal-frequency.
The SNR  is defined as 
${\text{SNR}}=\frac{3V_a^2}{\sigma^2}$.
The voltage magnitudes and phases are
considered to be  $V_a=1$,
$V_c=\beta V_a$
 per unit (p.u.), 
 $\varphi_a=\frac{\pi}{4}$ and  $\varphi_c=\varphi_a+\frac{2\pi}{3}+\epsilon$.
 For an almost balanced system,  we set $V_b=1.03 V_a$, $\varphi_b=\varphi_a-\frac{2\pi}{3}-\frac{3}{100}\pi$, $\beta=1$, and $\epsilon=0$.
A single-phase voltage magnitude and angle imbalance is implemented  by setting
 $\beta>1.03$ and $|\epsilon|>\frac{3}{100}\pi$.
The authorized level of imbalances of the  GLRT
is chosen to be $r=0.03$.
\subsection{Frequency estimation performance}
\begin{figure}
\label{Fig_freq}
      \begin{center}
			\vspace{-0.5cm}
         \begin{tabular}[t]{cc}
           \vspace{-0.75cm}\subfigure[$\Delta=0.1\times 2 \pi$]{ \vspace{-0.5cm}\centerline{\psfig{figure=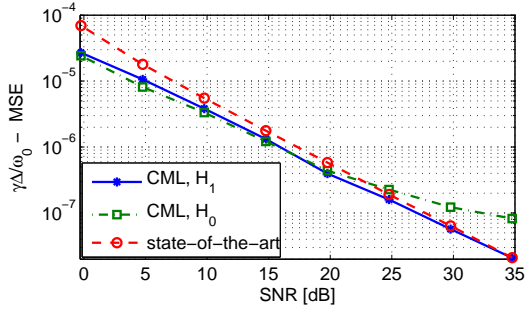,width=7.5cm}}\vspace{-0.75cm}}
\\
           \vspace{-0.5cm}\subfigure[$\Delta=2.5\times 2 \pi$]{ \vspace{-0.5cm}
					\centerline{\psfig{figure=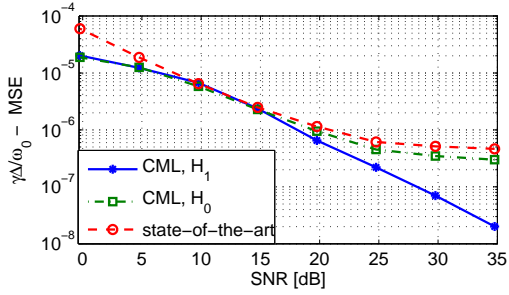,width=7.5cm}}\vspace{-0.25cm} }
        \end{tabular}
       \end{center}
			 \caption{The MSE of the normalized frequency deviation, $\gamma\frac{\Delta}{\omega_0}$, estimators for
 $K=12$, $N=48$,    and for
(a) $\Delta=0.1\times 2 \pi$; and
(b) $\Delta=2.5\times 2 \pi$.
			 }
			\label{Fig_freq_2}
      \end{figure}

The estimation performance of the normalized frequency deviation, $\frac{\gamma{\Delta}}{\omega_0}$,  is
evaluated  for an imbalanced model with
 $\beta=3$ and $\epsilon=0$.
The mean-square-error (MSE) of the state-of-the-art  frequency-estimator from (\ref{method1})
and the
CML estimators from (\ref{nice1s}),
under both $H_0$ and $H_1$,
are presented in Figs. \ref{Fig_freq_2}.a and \ref{Fig_freq_2}.b for $\Delta=0.1\times 2\pi $ and $\Delta=2.5\times 2\pi $, respectively.
It can be seen that for low SNR,  the CML estimators  under $H_0$ and $H_1$ 
 perform well and have similar performances for both frequencies.
However,  for high SNR, the CML estimator that assumes unbalanced system is significantly
 better than the CML estimator which assumes balanced system. 
The MSE of the CML estimator under $H_1$, i.e. under the unbalanced system assumption, is the lowest for any SNR. However, the CML estimators   suffer from high complexity and are affected by the search  resolution.
It can be seen that 
the state-of-the-art  frequency-estimator from (\ref{method1}) performs well for small frequency deviations, which is the
typical scenario in real-world power systems \cite{freq_deviation}.
For higher frequency deviations we derived in \cite{ICASSP_unbalanced} a low-complexity frequency estimation method for unbalanced system, which is beyond of the scope of this paper.

\subsection{Single-phase magnitude and phase imbalances}
\label{main_simulation}
The performance of the proposed GLRT
is compared  with the performance of the
  commonly-used   VUF method for detecting  voltage imbalance \cite{Jouanne_Banerjee_2001}, \cite{unbalanced_def}, \cite{unbalanced_defs}.
The  VUF test is defined as the ratio of the negative-sequence 
voltage magnitude 
 to the positive-sequence voltage magnitude.
In order to make a fair comparison,
we use the VUF definition  with 
 $K$  phasor  measurements of the positive- and negative-sequence:
\be
\label{TUVs}
T_{\text{VUF}}=\frac{\frac{1}{K}\sum_{k=0}^{K-1} \left|V_-[k]\right|}
{\frac{1}{K}\sum_{k=0}^{K-1} \left|V_+[k]\right|},
\ee
 which  is based only on the voltage magnitudes.
It should be noted that 
there is no analytical procedure for setting  the threshold of the VUF detector, $T_{\text{VUF}}$. 
In the following, we chose the threshold to maximize the probability of detection for each scenario.

In addition, in order to demonstrate power loss due to the unknown frequency deviation, i.e. reduction in detection probability for a given probability of error ratio, we also compare the results with a GLRT for known frequency deviation, $\Delta$.  This detector is given by the GLRT in 
(\ref{GLRT_pn}), in which we substitute  the known frequency deviation in
$P$, $Q$, $\evec_1$, and $\evec_2$, which affects the ML phasor estimators and $\kappa_1,\kappa_2$.
When the estimation error is small, the known frequency-deviation GLRT is expected to be close to the proposed GLRT.

\begin{figure}
\label{Fig7}
      \begin{center}
			\vspace{-0.5cm}
         \begin{tabular}[t]{cc}
           \vspace{-0.75cm}\subfigure[Magnitude imbalance]{ \vspace{-0.5cm}\centerline{\psfig{figure=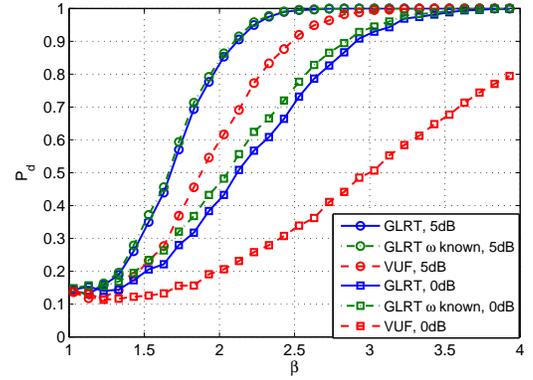,width=7cm}}\vspace{-0.75cm}}
\\
           \vspace{-0.5cm}\subfigure[Phase imbalance]{ \vspace{-0.5cm}\centerline{\psfig{figure=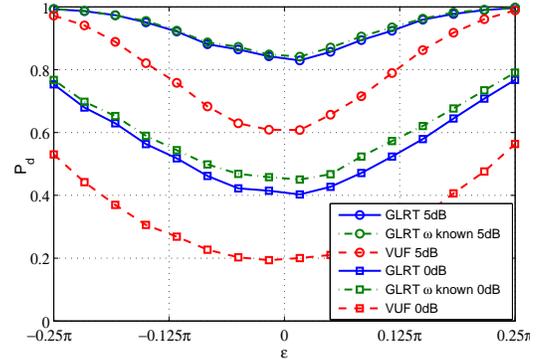,width=7cm}}\vspace{-0.25cm} }
        \end{tabular}
       \end{center}
			\caption{Magnitude and phase imbalances:
The  probability of detection of  VUF and GLRT with known/unknown frequency deviation
 for a constant probability of false alarm of $15\%$ and
 $K=12$, $N=48$, $\Delta=0.1\times 2 \pi$,  $V_a=1$, $\varphi_a=0.25\pi$ and
 SNR$=0,5$ dB are presented:
(a) versus  $\beta$  for $\epsilon=0$; and
(b) versus  $\epsilon$ for $\beta=2$.
			}
			\label{Fig8}
      \end{figure}

In this case, single-phase voltage magnitude and phase
imbalance is considered by changing the voltage
magnitude and phase of the single-phase $c$, i.e. by changing $\beta$ and $\epsilon$. 
In
	Figs. \ref{Fig8}.a and \ref{Fig8}.b, the probability of detection is presented versus different values of $\beta$ and $\epsilon$, respectively, for
	a constant false alarm probability of $15\%$.
	When $\beta$ approaches $1.03$ or $\epsilon$ approaches $\frac{3}{100}\pi$,  a reduction occurs  in the detection probability
	since the magnitude or phase voltage imbalance is smaller and identical to the balanced scenario.
	In this case, the probability of detection is equal to  the probability of error, i.e. equal to $0.15$.
	It can be seen that the detection probability of the GLRT is significantly higher than that of the VUF for any scenario.
	In addition,  for high SNR
	the performance of the GLRT with estimated frequency deviation coincides with the known-frequency GLRT.
It can be seen that the GLRT and VUF are robust to this scenario outside the local region of small insignificant imbalances and 
	these detectors are able to  distinguish between true imbalances (higher than $3\%$) and low unbalances.
	The  GLRT detection probability  is  higher than that of the VUF in this case too.
Fig. \ref{Fig8}.b examines that the proposed methods are symmetric w.r.t.
 clockwise and  anticlockwise movement of phasor
$c$.

\begin{figure}[htb]
\hspace{-0.25cm}
\centerline{\psfig{figure=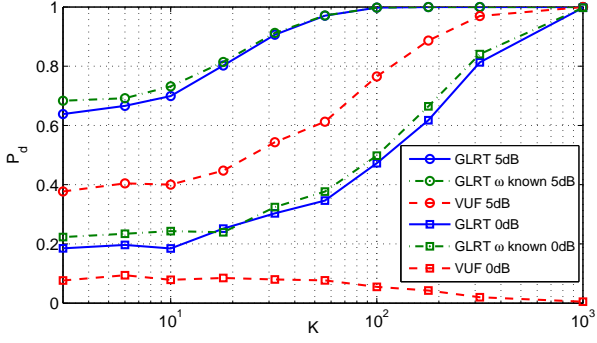,width=8.5cm}}
 \caption {
The  probability of detection of  VUF and GLRT with known/unknown frequency deviation
versus $K$
 for a constant probability of false alarm of $15\%$ and $\epsilon=0$,  $\beta=2$,
  $N=24$, $\Delta=0.1\times 2 \pi$,  $V_a=1$, $\varphi_a=0.25\pi$ and
 SNR$=0,5$ dB.
 }
 \label{Fig_K}
\end{figure}

In order to examine the influence of the number of  samples at the frequency domain, $K$, on the detection performance,  
the probability of detection is presented versus $K$ in \ref{Fig_K} for
	a constant false alarm probability of $15\%$ and  SNR$=0,5$ dB.
It can be seen that  for a large number of frequency domain samples, $K$, the effect of frequency deviation is reduced.
In particular,
for $K>15$, the performance of the  GLRT with estimated frequency deviation is very close to the performance of the known-frequency GLRT
for both SNR.
In contrast to the GLRTs, for
 low SNR an increase in   $K$ does not improve the performance since the VUF is not robust to the local-imbalances scenario. 
For high SNR, the probability of detection of the VUF increases with $K$ but it is lower than the probability of detection of the two GLRTs.

\subsection{Case study: Imbalance detection in the presence of higher order harmonics}
\begin{figure}
\label{Fig7_harmonics}
      \begin{center}
			\vspace{-0.5cm}
         \begin{tabular}[t]{cc}
           \vspace{-0.75cm}\subfigure[Magnitude imbalance]{ \vspace{-0.5cm}\centerline{\psfig{figure=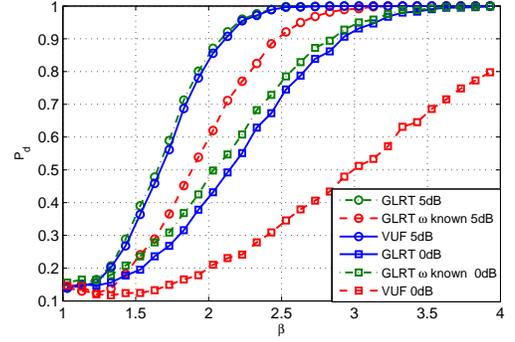,width=7cm}}\vspace{-0.75cm}}
\\
           \vspace{-0.5cm}\subfigure[Phase imbalance]{ \vspace{-0.5cm}\centerline{\psfig{figure=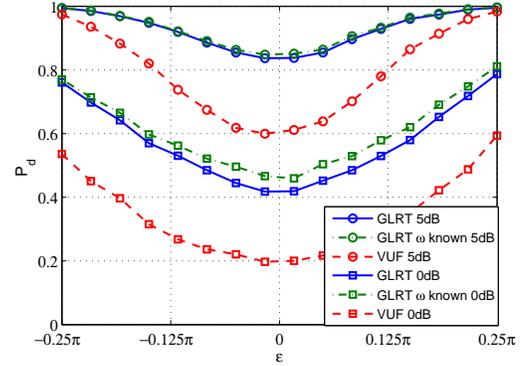,width=7cm}}\vspace{-0.25cm} }
        \end{tabular}
       \end{center}
			\caption{Magnitude and phase imbalances for non-sinusoidal signals:
The  probability of detection of  VUF and GLRT with known/unknown frequency deviation
 for a constant probability of false alarm of $15\%$ and
 $K=12$, $N=48$, $\Delta=0.1\times 2 \pi$,  $V_a=1$, $\varphi_a=0.25\pi$ and
 SNR$=0,5$ dB and with two additional harmonics are presented:
(a) versus  $\beta$  for $\epsilon=0$; and
(b) versus  $\epsilon$ for $\beta=2$.
			}
			\label{Fig8_harmonics}
      \end{figure}

Usually, there are additional harmonics 
where the harmonics frequencies
are multiples of the prevailing off-nominal network frequency \cite{PMUbook}.
The performance of the GLRT  is
 influenced by the 
imbalance degree and the voltage signal harmonic distortion.
In order to model
the influence of the harmonics, we replace
the model in (\ref{phasor_model2_three}) by (e.g. \cite{PMUbook}):
\beqna
\label{phasor_model2_three_generic}
\left[\begin{array}{c}v_a[n]\\v_b[n]\\v_c[n]
\end{array}
\right]
=\left[ \begin{array}{c} V_{a}\sum_{p=1}^P a_{p} \cos\left(
p(\gamma \frac{\omega_0+\Delta }{\omega_0}n+\varphi_a)\right) \\
V_b\sum_{p=1}^P a_{p}\cos\left(p(
\gamma \frac{\omega_0+\Delta }{\omega_0}n+\varphi_b)\right)\\
V_c\sum_{p=1}^P a_{p}\cos\left(p(
\gamma \frac{\omega_0+\Delta }{\omega_0}n+\varphi_c)\right)\end{array}
\right]\nonumber\\+\wvec_{a,b,c}[n],\hspace{4.5cm}\nonumber
\eeqna
where we set $P=4$, $a_{1}=1$, $a_2=0.2$, $a_3=0$, and $a_4=0.5$.
The other parameters are chosen to be the same  parameters as in Subsection \ref{main_simulation}.
The detectors' performance is presented for the case of non-sinusoidal and  voltage imbalance
 in Fig. \ref{Fig8_harmonics}.
By comparing the probabilities of detection in \ref{Fig8_harmonics} and \ref{Fig8}
it can be seen that the performance  degradation is not significant
and
 the proposed GLRT methods, as well as the existing VUF method, is not 
sensitive to inter-harmonics. Therefore, the proposed methods for detection of unbalances can be used also in the presence of harmonics.
\subsection{Probability of error}
\begin{figure}[htb]
\hspace{-0.25cm}
\centerline{\psfig{figure=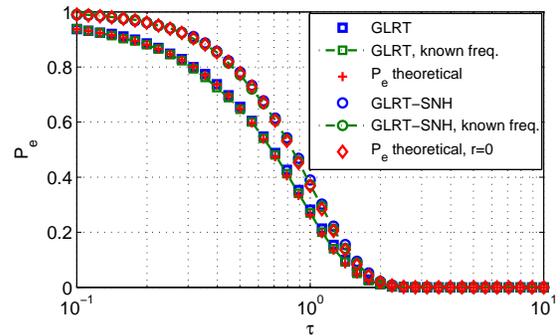,width=8cm}}
 \caption {The asymptotic and simulated probability of error of the GLRT
and
GLRT-SNH with known/unknown frequency deviation for
 $K=12$, $N=48$, $\Delta=0.1\times 2 \pi$,  $V_a=1$, $\varphi_a=0.25\pi$,
 $\beta=1.75$, $\epsilon=0.1\pi$, and
 SNR$=0$ dB.
 }
 \label{Fig6}
\end{figure}

The  simulated false alarm probability  of the GLRT and GLRT-SNH with known/unknown frequency deviation
	and the theoretical probability of error from (\ref{GLRT_Pe}) are presented in Fig. \ref{Fig6}
		versus the threshold, $\tau$,
	for 
an unbalanced  system with 
 $\beta=2$, $\epsilon=0.1\pi$,
$K=12$  frequency samples,
 and for an SNR of
 $10$ dB. 
	According to (\ref{GLRT_Pe}), the probability of error is a function of the threshold $\tau$
and $r$, but is independent of the noise level.
Therefore, Fig. \ref{Fig6} represents the results for any SNR.
It can be seen that  the theoretical false alarm probability  captures the behavior of the actual 
	false alarm probability of the GLRT and GLRT-SNH even in the unknown-frequency case. That is, 
 the theoretical asymptotic performance (or equivalently, the performance in the known frequency-deviation case)
adequately summarizes the actual performance for data records as short as $K=12$
samples. Thus, we can conclude that
although the asymptotic bound
theoretically requires an infinite number of observations, it
still provides a tight lower bound on the probability of error when there is a sufficiently large
observation window for both GLRT and GLRT-SNH. In addition, it can be verified that for the same threshold value, the GLRT-SNH has higher probability of error.
\section{Conclusion}
\label{diss}
In this study we demonstrate the detection of imbalances by using the PMU output of the symmetrical components when the voltage measurements are noise contaminated. 
We formulate the detection of imbalance  as a hypothesis testing problem with unknown constrained parameters within  the framework of detection theory.  The GLRT 
is derived for this  problem and the CML phasors' estimators are developed for both balanced and unbalanced systems and can be used for general state-estimation in a smart grid. 
The known-frequency and asymptotic performance 
of the proposed GLRT detector has been provided  and can be used as a benchmark. 
In detection theory, different tests induce different thresholds that the likelihood ratio is compared to  \cite{Kay_detection}. Thus, the  threshold setting is the key component of the hypothesis testing. In this context, a new formulation is devised for setting the threshold that also interpolates  the authorized level of imbalances.

Simulation results have verified that the proposed GLRT with
either known or estimated frequency deviation  yields competitive performance, compared to
the state-of-the-art VUF method and is better for magnitude imbalance detection.
In addition, we demonstrate that the proposed method is not sensitive to additional harmonics.
Topics for future research include the derivation of 
mitigation techniques that use the proposed GLRT as an imbalance measure in order to correct unbalanced voltage  problems more efficiently.
In addition, 
 real-time implementation of the proposed detectors can be very important, especially the  derivation of a change detection method for  imbalances.

\end{document}